\documentstyle[aps,prl,multicol,amssymb]{revtex}
\title{High order correlations of generic pure states of finite-dimensional
quantum systems are determined by lower order correlations}

\author{ N.~Linden$^1$ and W. K. Wootters$^2$}

\address{
$^1$School of Mathematics, University of Bristol, University
Walk, Bristol BS8 1TW, UK\\
$^2$Department of Physics, Williams College, Williamstown MA
01267, USA}

\date{13 August 2002}
\begin{document}
\maketitle
\begin{abstract}
We show that almost every pure state of multi-party quantum
systems (each of whose local Hilbert space has the same dimension)
is completely determined by the state's reduced density matrices
of a fraction of the parties; this fraction is less than about
two-thirds of the parties for states of large numbers of parties.
In other words once the reduced states of this
fraction of the parties have been specified, 
there is no further freedom in the state.

\end{abstract}


\newcommand{\mtx}[2]{\left(\begin{array}{#1}#2\end{array}\right)}
\newcommand{\tr}{\mbox{Tr} }
\newcommand{\ket}[1]{\left | #1 \right \rangle}
\newcommand{\bra}[1]{\left \langle #1 \right |}
\newcommand{\amp}[2]{\left \langle #1 \left | #2 \right. \right \rangle}
\newcommand{\proj}[1]{\ket{#1} \! \bra{#1}}
\newcommand{\ave}[1]{\left \langle #1 \right \rangle}
\newcommand{\superop}{{\cal E}}
\newcommand{\unity}{\mbox{\bf I}}
\newcommand{\hilbert}{{\cal H}}
\newcommand{\relent}[2]{S \left ( #1 || #2 \right )}
\newcommand{\banner}[1]{\bigskip \noindent {\bf #1} \medskip}

\begin{multicols}{2}
It is natural to think that a pure quantum state of $n$ parties,
chosen at random, would have some multi-party entanglement of all
possible types including what one might call irreducible $n$-party
entanglement.  Giving concrete, and quantifiable, meaning to this
idea is a major goal in the foundations of quantum mechanics 
and quantum information
theory which has yet to be achieved.

Nonetheless, it has been shown that not all entanglement of
$n$-parties can be reversibly transformed into two-party
entanglement \cite{Bennett99,LPSW,Acin}, and indeed
\cite{Bennett99,LPSW} that for any $n$ there are states which
cannot be transformed reversibly into states of fewer than $n$
parties.  One might deduce from this that there is a notion of
irreducible $n$-party entanglement, even though it has so far
eluded us as to how to measure the amount of it that is contained
in any given $n$-party state.  We note that the general situation
is different from the case of two parties where the entropy of
entanglement is essentially the unique measure of the bipartite
entanglement of two-party pure states \cite{BBPS}.

In this letter we give results which throw a surprising light on
these issues.  We consider the  case of pure states of any number
$n$ of parties each of which has a $d$-dimensional Hilbert space.
We show that for almost all such states ({\em i.e.} for generic states
of this type), the reduced states of a fraction of the parties (at
most about two-thirds for large $n$)
 uniquely specify the full quantum state of the
$n$ parties; there are no other states, {\em pure or mixed}
consistent with the given reduced states. In the language of
\cite{LPW}, we may say that all the information in a generic 
$n$-party 
state is contained in the reduced states of a fraction of
the parties. Expressed differently, the low order correlations
uniquely determine the high order correlations.

An earlier paper \cite{LPW} considered this question for pure
states of three qubits.  It was shown  that in this case the three
two-party reduced states uniquely determine the full three-party
state for generic pure states of three parties.  It might have
been imagined that this was an anomalous case arising from the low
dimensionality of the system.  We show here that, on the contrary,
this general picture, namely that the high order correlations are
determined by lower order ones, is the rule for generic pure
quantum states in finite dimensions.

It may be worth remarking how different this situation is from the 
case of classical probability distributions.  For generic 
distributions of $n$ random variables each taking $d$ values---such 
distributions arise, for example, from making local von Neumann 
measurements on the quantum systems we are considering---it is not 
difficult to show that even the set of 
all the marginal distributions 
for $n-1$ of the variables does not uniquely specify the full 
probability distribution.

In a different context---the many-electron systems of molecular
physics---much progress has been made on the related problem of
reconstructing $n$-electron density matrices (especially where
$n=3$ or 4) from the 2-particle reduced state
\cite{reconstruct,book}. Although our problem is similar in
spirit, in molecular physics it is fundamental that the particles
are indistinguishable (fermions) and so the fact that full quantum
state is totally antisymmetric plays a key role. In this letter,
we deal with the usual context of quantum information theory: the
particles are distinguishable and the full quantum state need have
no particular symmetry under interchange of the particles.

The plan of this letter is first to show there is a fraction
$\alpha_U$ of the parties such that, given the reduced states of
this fraction of the parties, the only state (pure or mixed)
consistent with these reduced states is the original state (the
subscript $U$ is to denote the fact that this is an upper bound).
These reduced states uniquely specify the state and all the
information in the full state is already contained in the reduced
states. This result, however, leaves open the possibility that
perhaps the true proportion of parties whose reduced states
uniquely specify the full state is much smaller; indeed it might
grow like $\log n$ for example.  The second part of the letter
shows that, in fact, the number of parties must grow linearly with
$n$.  We give a lower bound $\alpha_L$ for this proportion; it is
about $18.9\%$ for systems of $n$ qubits and grows to 50\% for
systems of $n$ $d$-level systems.

We will first show that there is an upper bound $\alpha_U$ 
on the fraction
of parties whose reduced states are sufficient to
uniquely specify the full state of $n$ parties.  To this end we
first consider three parties $A,B,C$ with Hilbert spaces whose
dimensions are $M,N,P$ respectively.  We will take $M\geq N+P-1$
for reasons which will become clear.

Consider then, an arbitrary pure state $|\xi\rangle$ of three
parties $A$, $B$ and $C$.  We can write $|\xi\rangle$ as
\begin{equation}
|\xi\rangle = \sum_{i=1}^M \sum_{j=1}^N\sum_{k=1}^P
a_{ijk}|ijk\rangle;
\end{equation}
the labels in the ket refer to systems $A$, $B$ and $C$ in that
order. We wish first to answer the following question: Under what
conditions is $|\xi\rangle$ uniquely determined by its
two-particle reduced states?

A state that agrees with $|\xi\rangle$ in its reduced states but
is not equal to $|\xi\rangle$ is most likely going to be a mixed
state. In order to allow for this possibility, it is helpful to
imagine an environment $E$ with which the system might be
entangled, such that the whole system, system plus environment, is
in a pure state $|\psi\rangle$. Let us first ask what form
$|\psi\rangle$ must take in order to be consistent with the
(generally mixed) state of the pair $AB$ derived from
$|\xi\rangle$.

The density matrix of this two-particle state is at most of rank
$P$, being confined to the space spanned by the vectors
$|v_1\rangle = \sum_{ij} a_{ij1}|ij1\rangle,\ |v_2\rangle =
\sum_{ij} a_{ij2}|ij2\rangle,\ldots , |v_P\rangle = \sum_{ij}
a_{ijP}|ijP\rangle$.  The state $|\psi\rangle$ must thus be a
superposition of the form
 \begin{equation}
|\psi\rangle = |v_1\rangle|E_1\rangle
 + |v_2\rangle |E_2\rangle\ldots +|v_P\rangle|E_P\rangle ,
 \end{equation}
$|E_1\rangle,\ldots ,|E_P\rangle$ being states of the joint system
$CE$.  Moreover, if the states $|v_1\rangle,
\ldots, |v_P\rangle$
are linearly independent---this will be the case for almost all
states $|\xi\rangle$---then in order to get the
correct density matrix when
one traces over $C$ and $E$, the $P$ states $|E_1\rangle,\ldots
,|E_P\rangle$ must be orthonormal. Expanding $|E_1\rangle,\ldots
,|E_P\rangle$ in the standard basis $\{|1\rangle,\ldots, |P\rangle
\}$ of particle $C$, we obtain the following form for
$|\psi\rangle$:
\begin{equation}
|\psi\rangle = \sum_{i=1}^M \sum_{j=1}^N\sum_{k,l=1}^P
a_{ijl}|ijk\rangle |e_{lk}\rangle. \label{formAB}
\end{equation}
The states $|e_{lk}\rangle$ are states of $E$ alone.  The
orthonormality conditions on $|E_1\rangle, \ldots,|E_P\rangle$
become
\begin{equation}
\sum_k \langle e_{lk}|e_{l'k}\rangle = \delta_{ll'}.
\label{orthAB}
\end{equation}
Thus, in order to match the reduced state
on $AB$, $|\psi\rangle$ must
be of the form given in Eq.~(\ref{formAB}), and for a
generic $|\xi\rangle$, the $|e_{lk}\rangle$ in this equation
must satisfy Eq.~(\ref{orthAB}).

Similarly, in order to match the reduced state
on $AC$, we must have
\begin{equation}
|\psi\rangle = \sum_{i=1}^M \sum_{j,l=1}^N\sum_{k=1}^P
a_{ilk}|ijk\rangle |f_{lj}\rangle, \label{formAC}
\end{equation}
where the states $|f_{lj}\rangle$ are states of $E$ satisfying
(for generic $|\xi\rangle$)
\begin{equation}
\sum_j \langle f_{lj}|f_{l'j}\rangle = \delta_{ll'}.
\label{orthAC}
\end{equation}
And to match the reduced state on $BC$, we must have
\begin{equation}
|\psi\rangle = \sum_{i,l=1}^M \sum_{j=1}^N\sum_{k=1}^P
a_{ljk}|ijk\rangle |g_{li}\rangle, \label{formBC}
\end{equation}
with (again for generic $|\xi\rangle$)
\begin{equation}
\sum_i \langle g_{li}|g_{l'i}\rangle = \delta_{ll'}.
\label{orthBC}
\end{equation}
We will proceed by  deriving consequences of the two equations
(\ref{formAB}) and (\ref{formAC})---both of these expressions
must describe the same state $|\psi\rangle$.  Thus

\begin{eqnarray}
& &\sum_{i=1}^M \sum_{j=1}^N\sum_{k,l=1}^P a_{ijl}|ijk\rangle
|e_{lk}\rangle\nonumber\\
& &\quad =\sum_{i=1}^M \sum_{j,r=1}^N\sum_{k=1}^P
a_{irk}|ijk\rangle |f_{rj}\rangle.
\end{eqnarray}

We now consider specific terms in this equation. For example
consider the terms with $|i11\rangle_S$ in them with $i$ fixed.
They lead to $M$ equations (one for each choice of $i$)
\begin{equation}
\sum_{l=1}^P a_{i1l} |e_{l1}\rangle = \sum_{r=1}^N a_{ir1}
|f_{r1}\rangle . \label{Mequations11}
\end{equation}

It is helpful to rearrange these equations as $M$ homogeneous
equations in the $N+P-1$ variables
\begin{equation}
(|e_{11}\rangle-|f_{11}\rangle),
|e_{21}\rangle,\ldots|e_{P1}\rangle,|f_{21}\rangle,\ldots|f_{N1}\rangle .
\end{equation}

Let us take the case that $M\geq N+P-1$.  In this case, for
generic values of the $a_{ijk}$, the only solutions are
\begin{equation}
 |e_{11}\rangle = |f_{11}\rangle;\quad |e_{l1}\rangle =0\  {\rm for}\ l\neq 1
 ;\quad |f_{r1}\rangle =0\  {\rm for}\ r\neq 1 .
\end{equation}
Note that the $M$ equations do not involve all the $a_{ijk}$, and
there is no reason for the associated determinant to be zero in
general.

Now consider the equations with with $|i12\rangle$ in them with
$i$ fixed.   These are $M$ equations
\begin{equation}
\sum_{l=1}^P a_{i1l} |e_{l2}\rangle = \sum_{r=1}^N a_{ir2}
|f_{r1}\rangle . \label{Mequations12}
\end{equation}
Using the fact that the only non-zero $|f_{r1}\rangle$ is
$|f_{11}\rangle$, we can rearrange (\ref{Mequations12}) into $M$
equations in the $P$ variables
\begin{equation}
(|e_{22}\rangle-|f_{11}\rangle),
|e_{12}\rangle,|e_{32}\rangle,\ldots|e_{P2}\rangle.
\end{equation}
These equations will have solutions
\begin{equation}
 |e_{22}\rangle = |f_{11}\rangle;\quad |e_{l2}\rangle =0\  {\rm for}\ l\neq
 2
,
\end{equation}
since again the determinant will not be zero, in the generic case.

Proceeding in this way to use the equations in $|i1k\rangle_S$,
$k=1\ldots P$, we find eventually that
\begin{equation}
 |e_{lk}\rangle = \delta_{lk}| e_{11} \rangle.
\end{equation}
Thus
\begin{equation}
|\psi\rangle = \sum_{ijkl} a_{ijl}|ijk\rangle |e_{lk}\rangle =
\sum_{ijk} a_{ijk}|ijk\rangle |e_{11}\rangle.
\end{equation}

In other words, the fact that $|\psi\rangle$ must be consistent
with the reduced states of $AB$ and $AC$ forces it to be the
original pure state tensor product with a state of the
environment.  We note that, in getting to this result, we have not
needed to make use of the requirement that the full state be
consistent with the reduced state for $BC$.

We may now use this three-party result to learn about $n$-party
systems.  For let $N=P=d^m$, $M=d^{(m+1)}$, {\em i.e.}, a total of
$(3m+1)$ $d$-level systems (clearly $M>N+P-1$).  The reduced
states of $(2m+1)$ parties, determine the full state.  In other
words, for large numbers of parties, the knowledge of the reduce
states of roughly $\alpha_U=2/3$ of the parties is sufficient to
uniquely specify the full pure state.

We notice that we have made no use of the orthogonality conditions
for the environment states. Indeed in the case of three qubits, we
were able to use this orthogonality to show that the three
two-party reduced states uniquely specific the full state of three
parties for generic pure 3-qubit states. We thus expect that the
orthogonality conditions will allow us to reduce the fraction
$\alpha_U$ of parties whose reduced states are required to specify
the full state.

It will also have been noticed that we have derived the above
bound for $n$-party systems by requiring that the full state be
consistent with only two of the very many reduced states of full
system.  One may well imagine, that requiring consistency with the
all the reduced states reduces this fraction.  Perhaps the number
of parties needed might be much less than 
two-thirds of the total. Indeed
perhaps it might grow sub-linearly with  $n$.  We now show that in
fact the true number of parties cannot be much less than our upper
bound $n\alpha_U$: the number must grow linearly with $n$ and
indeed for $n$ qubits it must be more than about 0.189$n$ for
large $n$. We do this by finding a lower bound $\alpha_L$; one
must know the reduced states of at least this fraction of the
parties.

Let us first consider the case of qubits.  We consider a fraction
$\alpha$ of the total number of qubits $n$. We have in mind that
all the reduced states of this fraction of qubits are known.  The
total number of parameters in this set of reduced states must
certainly be as large at the number $2^{n+1}-2$ of parameters in
the pure states we could hope to reconstruct.

We now estimate how many independent parameters there are in the
reduced states of $n\alpha$ of the particles.  We use the Bloch
parametrization which is valid for any state, pure as well as
mixed, {\em e.g.} for 3 parties any density matrix may be written as
\begin{eqnarray}
& &\rho_{ABC} ={1\over 8}\Big( 1 \otimes 1\otimes 1 +\alpha_i
\sigma_i \otimes 1\otimes 1 + \beta_i 1 \otimes \sigma_i \otimes 1
\nonumber\\
& &\quad + \gamma_i 1 \otimes 1 \otimes \sigma_i + R_{ij} \sigma_i
\otimes \sigma_j \otimes 1 + S_{ij} \sigma_i \otimes 1\otimes
\sigma_j \nonumber \\
& &\quad + T_{ij} 1\otimes \sigma_i \otimes \sigma_j  + Q_{ijk}
\sigma_i \otimes \sigma_j \otimes \sigma_k \Big),\label{bloch}
\end{eqnarray}
since the set of matrices $(1,\sigma_x,\sigma_y,\sigma_z)$ is a
basis for the operators on ${\mathbb C}^2$ (the parameters
$\alpha_i,\beta_i$ etc. are real).

Thus for $n$ parties the total number of parameters in the reduced
states of up to $n\alpha$ parties is
\begin{equation}
\sum_{r=1}^{n\alpha} \mtx{c}{n \\ r}\ 3^r.
\end{equation}

Now
\begin{eqnarray}
{\mtx{c}{n\\ r-1}\ 3^{r-1}\over \mtx{c}{n \\ r}\ 3^r}
& & = {r\over 3(n-r+1)}\nonumber\\
& &\leq {\alpha\over 3(1-\alpha)},
\end{eqnarray}
for $r\leq n\alpha$.

Thus
\begin{eqnarray}
\sum_{r=1}^{n\alpha} \mtx{c}{n \\ r}\ 3^r\leq \mtx{c}{n \\
n\alpha}\ 3^{n\alpha} \left({3-3\alpha\over 3-4\alpha}\right),
\end{eqnarray}
summing the geometric progression to infinity (we need $\alpha<
3/4$, but it will be, see below).

Now we find a value of $\alpha$ for which this total number of
terms in the reduced states is less than $2^{n+1}-2$.  We want
\begin{equation}
\mtx{c}{n \\ n\alpha}\ 3^{n\alpha} \left({3-3\alpha\over
3-4\alpha}\right)\leq 2^{n+1}-2.
\end{equation}

Thus at leading order in $n$, we need
\begin{equation}
e^{nH(\alpha)+n\alpha \ln 3} \leq e^{n\ln 2},\label{finalineq}
\end{equation}
where $H(x) = -x \ln x - (1-x) \ln (1-x)$.

Numerically we find the solution to
\begin{equation}
H(\alpha)+\alpha \ln 3-\ln 2=0\label{qubit-condition}
\end{equation}
to be $\alpha \sim 0.189$.  Thus for $\alpha$ less than this there
are not enough parameters in the reduced states to account for the
different pure states.

Thus taking the upper and lower bounds together we conclude that
for generic pure states of $n$ qubits, the reduced states of
somewhere between $0.189n$ and $2n/3$ of the qubits  uniquely
specify the full quantum state; differently put the correlations
amongst between $0.189$ and $2/3$ of the qubits specify uniquely
the high order correlations ({\em i.e.} the correlations amongst more of
the qubits).  Using the analogue of the parametrization in
(\ref{bloch}) for an $n$-party state, we see that the higher order
tensors in the expression for the pure state are determined by the
lower order ones; these higher order tensors may not be freely
varied once the lower order tensors are specified.

For systems of $n$ parties, each of which lives in a
$d$-dimensional Hilbert space, the argument for the lower bound is
the same as that which we used for qubits. Rather than the 3 Pauli
matrices we use $d^2-1$ matrices to span the space of traceless
Hermitian operators. Thus the condition (\ref{qubit-condition})
becomes
\begin{equation}
H(\alpha)+\alpha \ln (d^2-1)-\ln d=0.\label{d-condition}
\end{equation}
One finds that the lower bound for the fraction increases with
increasing $d$; from 18.9\% for qubits to $1/2$ for large $d$.

We believe that it will be valuable to find the exact fraction of
parties whose reduced states determine the full state for general
values of $n$ and $d$.  The states which are {\em not} generic are
also interesting.  In the case of three qubits \cite{LPW}, the
(non-trivial) non-generic states are those which are locally
equivalent to
\begin{equation}
a |111\rangle + b|222\rangle.
\end{equation}
There are many states consistent with the two-party reduced states
for these pure states.  We conjecture that non-generic states for
general $n$ and $d$ will have special properties as far as their
multi-particle entanglement is concerned.

We thank Sandu Popescu for illuminating conversations. We
gratefully acknowledge funding from the European Union under the
project EQUIP (contract IST-1999-11063).

\end{multicols}

\begin{thebibliography}{10}
\bibitem{Bennett99}C.H.
Bennett, S. Popescu, D. Rohrlich, J.A. Smolin and A.V. Thapliyal,
Phys. Rev. A65 (2001) 012307.
\bibitem{LPSW} N. Linden, S. Popescu, B. Schumacher and M. Westmoreland,
quant-ph/9912039
\bibitem{Acin} A. Acin, G. Vidal and J.I. Cirac, quant-ph/0202056
\bibitem{BBPS} C. H. Bennett, H. J. Bernstein, S. Popescu, and B. Schumacher,
Phys. Rev. A {\bf 53}, 2046 (1996); S. Popescu and D. Rohrlich,
Phys. Rev. A {\bf 56}, R3319 (1997).
\bibitem{LPW} N. Linden, S. Popescu and W.K. Wootters,
quant-ph/0207109
\bibitem{reconstruct} See, for example, F. Colmenero, C. Perez del Valle
and C. Valdemoro, Phys. Rev. A {\bf 47}, 971 (1993);
F. Colmenero and C. Valdemoro, Phys. Rev. A {\bf 47}, 979 (1993);
H. Nakatsuji and K. Yasuda, Phys. Rev. Lett. {\bf 76}, 1039 (1996);
K. Yasuda and H. Nakatsuji, Phys. Rev. A {\bf 56}, 2648 (1997);
D. A. Mazziotti, Phys. Rev. A {\bf 57}, 4219 (1998);
D. A. Mazziotti, Phys. Rev. A {\bf 60}, 3618 (1999).
\bibitem{book} For a historical review, see A. J. Coleman and V. I. Yukalov,
{\em Reduced Density Matrices: Coulson's Challenge}
(Springer-Verlag, New York, 2000).




\end{thebibliography}
\end{document}